\input harvmac
\noblackbox
 
\font\ticp=cmcsc10
 
\def\Title#1#2{\rightline{#1}\ifx\answ\bigans\nopagenumbers\pageno0\vskip1in
\else\pageno1\vskip.8in\fi \centerline{\titlefont #2}\vskip .5in}

\font\ticp=cmcsc10
\font\ttsmall=cmtt10 at 8pt

\input epsf
\ifx\epsfbox\UnDeFiNeD\message{(NO epsf.tex, FIGURES WILL BE
IGNORED)}
\def\figin#1{\vskip2in}
\else\message{(FIGURES WILL BE INCLUDED)}\def\figin#1{#1}\fi
\def\ifig#1#2#3{\xdef#1{fig.~\the\figno}
\goodbreak\topinsert\figin{\centerline{#3}}%
\smallskip\centerline{\vbox{\baselineskip12pt
\advance\hsize by -1truein\noindent{\bf Fig.~\the\figno:} #2}}
\bigskip\endinsert\global\advance\figno by1}


\lref\fgl{B.~Freivogel, S.~B.~Giddings and M.~Lippert,
{\it Toward a theory of precursors},
hep-th/0207083.}
\lref\dkk{U. Danielsson, E. Keski-Vakkuri and M. Kruczenski, {\it
Spherically Collapsing Matter in AdS, Holography, and Shellons},
Nucl. Phys. B563 (1999) 279, hep-th/9905227.}
\lref\dkkk{U. Danielsson, E. Keski-Vakkuri and M. Kruczenski, {\it
Vacua, Propagators, and Holographic Probes in AdS/CFT},
JHEP 9901 (1999) 002, hep-th/9812007.}
\lref\dkkbh{
U.~H.~Danielsson, E.~Keski-Vakkuri and M.~Kruczenski,
{\it Black hole formation in AdS and thermalization on the boundary},
JHEP 0002, 039 (2000)
hep-th/9912209.}
\lref\simon{V. Balasubramanian and S. F. Ross,
{\it Holographic Particle Detection},
 Phys. Rev. D61 (2000) 044007, 
hep-th/9906226}
\lref\juan{J. Maldacena, 
{\it The Large N Limit of Superconformal Field Theories and Supergravity},
Adv. Theor. Math. Phys. 2 (1998) 231, hep-th/9711200.}
\lref\magoo{O. Aharony, S.S. Gubser, J. Maldacena, H. Ooguri, Y. Oz,
{\it Large N Field Theories, String Theory and Gravity},
Phys. Rept. 323 (2000) 183, hep-th/9905111.}
\lref\witten{ E. Witten, 
{\it Anti De Sitter Space And Holography},
Adv. Theor. Math. Phys. 2 (1998) 253, hep-th/9802150.}
\lref\witt{ E. Witten, 
{\it Anti-de Sitter Space, Thermal Phase Transition, 
And Confinement In Gauge Theories},
Adv. Theor. Math. Phys. 2 (1998) 505, hep-th/9803131.}
\lref\gkp{S. Gubser, I. Klebanov, and A. Polyakov, 
{\it Gauge Theory Correlators from Non-Critical String Theory},
Phys. Lett. B428 (1998) 105,
hep-th/9802109.}
\lref\hoit{G. Horowitz and N. Itzhaki, 
{\it Black Holes, Shock Waves, and Causality in the AdS/CFT Correspondence},
JHEP 9902 (1999) 010, hep-th/9901012}
\lref\gary{G. Horowitz, 
{\it Comments on Black Holes in String Theory},
Class. Quant. Grav. 17 (2000) 1107, hep-th/9910082.}
\lref\peet{A. Peet and J. Polchinski, 
{\it UV/IR Relations in AdS Dynamics},
Phys. Rev. D59 (1999) 065011, 
hep-th/9809022.} 
\lref\suwi{L. Susskind and E. Witten, 
{\it The Holographic Bound in Anti-de Sitter Space},
hep-th/9805114.} 
\lref\suto{L. Susskind and N. Toumbas,
{\it Wilson Loops as Precursors},
Phys.Rev. D61 (2000) 044001, hep-th/9909013 .}
\lref\pst{J. Polchinski, L. Susskind, and N. Toumbas,
{\it Negative Energy, Superluminosity and Holography},
Phys.Rev. D60 (1999) 084006, hep-th/990322.}
\lref\lmr{J. Louko, D. Marolf, and S. F. Ross,
{\it On geodesic propagators and black hole holography},
Phys.Rev. D62 (2000) 044041, hep-th/0002111.}
\lref\giro{S.  Giddings and S. Ross,
{\it D3-brane shells to black branes on the Coulomb branch},
Phys.Rev. D61 (2000) 024036,
hep-th/9907204.}
\lref\gili{S.~B.~Giddings and M.~Lippert,
{\it Precursors, black holes, and a locality bound},
Phys.\ Rev.\ D 65, 024006 (2002),
hep-th/0103231.}
\lref\wl{D.~Beckman, D.~Gottesman, A.~Kitaev and J.~Preskill,
{\it Measurability of Wilson loop operators},
Phys.\ Rev.\ D 65, 065022 (2002),
hep-th/0110205.}
\lref\so{G.~T.~Horowitz and V.~E.~Hubeny,
{\it CFT description of small objects in AdS},
JHEP 0010, 027 (2000),
hep-th/0009051.}
\lref\matschull{
H.~J.~Matschull,
{\it Black hole creation in 2+1-dimensions},
Class.\ Quant.\ Grav.\  16, 1069 (1999),
gr-qc/9809087;
S.~Holst and H.~J.~Matschull,
{\it The anti-de Sitter Gott universe: A rotating BTZ wormhole},
Class.\ Quant.\ Grav.\  16, 3095 (1999)
gr-qc/9905030.}
\lref\loma{
J.~Louko and D.~Marolf,
{\it Single-exterior black holes and the AdS-CFT conjecture},
Phys.\ Rev.\ D 59, 066002 (1999),
hep-th/9808081.}
\lref\maldacena{J.~M.~Maldacena,
{\it Eternal black holes in Anti-de-Sitter},
hep-th/0106112.}
\lref\bklt{
V.~Balasubramanian, P.~Kraus, A.~E.~Lawrence and S.~P.~Trivedi,
{\it Holographic probes of anti-de Sitter space-times},
Phys.\ Rev.\ D 59, 104021 (1999),
hep-th/9808017.}
\lref\gregross{
J.~P.~Gregory and S.~F.~Ross,
{\it Looking for event horizons using UV/IR relations},
Phys.\ Rev.\ D 63, 104023 (2001),
hep-th/0012135.}
\lref\kali{
D.~Kabat and G.~Lifschytz,
{\it Gauge theory origins of supergravity causal structure},
JHEP 9905, 005 (1999), hep-th/9902073.}
\lref\bdhm{
T.~Banks, M.~R.~Douglas, G.~T.~Horowitz and E.~J.~Martinec,
{\it AdS dynamics from conformal field theory},
hep-th/9808016.}

%
\baselineskip 16pt
\Title{\vbox{\baselineskip12pt
\line{\hfil   SU-ITP-2/32}
\line{\hfil \tt hep-th/0208047} }}
{\vbox{
{\centerline{Precursors see inside black holes}
}}}
\centerline{\ticp Veronika E. Hubeny\footnote{}{\ttsmall
 veronika@itp.stanford.edu}}
\bigskip
\centerline{\it Department of Physics, Stanford University,
Stanford, CA 94305, USA}
\bigskip
\centerline{\bf Abstract}
\bigskip
We argue that, given the nonlocal nature of precursors in AdS/CFT 
correspondence,
the boundary field theory contains information about events
inside a black hole horizon.
The essence of our proposal is sketched in figure 1, and relies
on the global nature of event horizons.

\Date{August, 2002}
\newsec{Introduction}

With the pressing problems related to string theory in cosmological 
settings, there has been a resurgence of interest in physics associated
with interiors of black holes.
So far, the best handle we have appears in the context of the AdS/CFT 
correspondence \refs{\juan,\witten,\gkp,\magoo}, 
where string theory on AdS (which is a gravitational
theory, and can contain black holes) has a dual description in terms of
a field theory ``living on the boundary''.
The details of the correspondence remain yet to be unraveled, and
its causal properties to be better understood; nevertheless, there are
certain expectations on how the bulk information is to be holographically
encoded on the boundary.

One system in which we can try to ask these questions is the large 
Schwarzschild black hole in $AdS_5 \times S^5$. 
This is well known to correspond to an approximately thermal state in the
gauge theory.
The spacetime has an event horizon, i.e.\ the black hole interior is causally
disconnected from the boundary.
One may then ask: {\it Is the physics inside the event horizon encoded by the 
boundary theory?}
Affirmative answer immediately leads to the
 more ambitious question\foot{
Such questions have been considered previously
\refs{\simon,\lmr,\gili,\maldacena}, but
despite extensive work, these issues still remain somewhat obscure.
The holographic representation of horizons in terms of the
dual gauge theory has also been discussed 
\refs{\bdhm,\bklt,\kali,\dkkbh,\gregross};
we believe that the lack of complete success in this venue 
stems largely from the global nature of the horizon.}: {\it How?}  
Once this is understood, one may hope
to come closer to the resolution of the black hole information paradox,
as well as to get a better handle on the singularity inside the horizon.  
While these are certainly worthy---and rather lofty---goals by themselves, 
a better
understanding and ``resolution'' of spacelike singularities would provide
a new window into studying cosmological singularities in string theory.

In this brief note, we point out that under the commonly-accepted assumption 
of precursors, at least part of the spacetime inside the horizon will be
accessible to the field theory.  
In the following we will explain this idea in more detail, using a
simple gedanken experiment which exemplifies it.  No explicit calculations
are presented, since it is the matter-of-principle point that we wish to 
stress.

\newsec{Gedanken experiment}

Let us, for the moment, consider pure AdS, without any black hole present.
Even certain local boundary operators contain a 
remarkable amount of information about static objects deep inside 
the bulk of AdS \so;
but since such operators are determined only from the asymptotic values of
the bulk fields,
any dynamical information is propagated out in a manner
 consistent with bulk causality.  
In particular, no local  operators on the boundary
can see events inside a causally disconnected region, such as a black hole.
However, as has been argued \refs{\pst,\suto,\fgl}, 
an event in the bulk is actually encoded by  
{\it nonlocal}
operators on the boundary, with the associated scale dictated by the UV/IR 
correspondence \refs{\suwi,\peet}.
A concrete proposal has been put forth that such operators are decorated 
Wilson loops.  
While this has not yet been proved\foot{
The identification of precursors with Wilson loops has been suggested in
\pst\ and expanded in \suto; while the latter's calculation was challenged 
by \gili,  further analysis \fgl\ suggested that the precursor
information is coded in high frequency components of Wilson loops, or
the so-called decorated Wilson loops.},
the essential point is the existence of {\it some} set
of nonlocal operators encoding the bulk, apparently ``instantaneously'', 
i.e.\ before any message sent from the event could reach the boundary.

If we accept this assumption, the fact that the boundary field theory 
must also contain some information about physics inside a horizon 
follows almost immediately.  
(A similar observation has been made recently in \gili, but their 
attempts at exemplifying the process through ``flossing the black hole''
cast some doubt on the possibility of extracting information  from inside
the horizon.
Here we suggest a much simpler process, which does not rely on 
 inconclusive tree level computations.)
The event horizon is a global object, defined as the boundary of the
past of the future infinity,
which means that we cannot determine the presence or position of the
event horizon until we know the entire future evolution of the spacetime.
In particular, an innocent-looking, empty, locally AdS part 
of spacetime can in fact be an interior of a black hole, if  later
there forms a large enough black hole for its horizon to stretch far enough
into the past to encompass the event in question.

\ifig\figa{Sketch of the proposed process of abstracting information 
from inside of the horizon: 
a) An event $p$, in locally pure AdS space is measured by $W$.
b) After this measurement is performed, a shell 
$s$ collapses, forming a black hole with a horizon $H$ which encompasses
$p$.}
{\epsfxsize=10cm \epsfysize=8cm \epsfbox{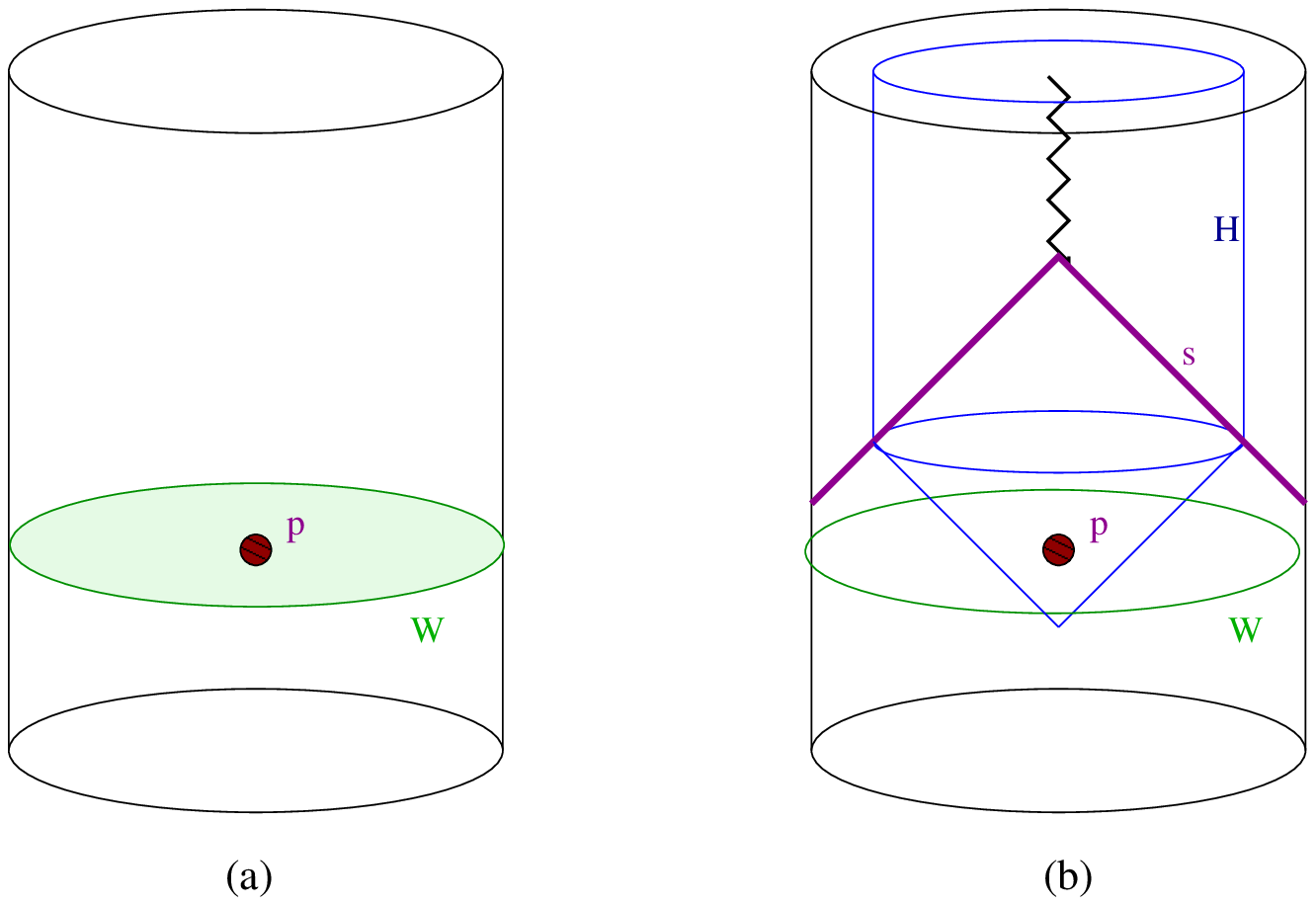}}

Taking advantage of this fact, we can consider the following gedanken
experiment:
Start with ``empty'' AdS space, 
and consider some event, labeled $p$ in Fig.1a. 
As a boundary observer, one can obtain instantaneous information
about the event $p$, for instance by measuring\foot{
While it was argued \wl\ that a Wilson loop cannot be measured in 
the conventional non-demolition sense, it can be measured in a 
suitable demolition experiment, such as sketched out in \fgl, which 
is all we need for the present argument.} 
appropriately decorated Wilson loop $W$.
{\it After} this measurement has been made, i.e.\ entirely to the 
future of $W$, one can send in a shell $s$ of radiation with sufficient
energy such that when it implodes in the center of AdS,
it forms a large black hole, as sketched in Fig.1b.
If the shell is spherically symmetric, the spacetime everywhere outside
this shell will be that of Schwarzschild-$AdS_5 \times S^5$, whereas 
inside the shell the spacetime is described by pure $AdS_5 \times S^5$.
The global event horizon for this spacetime will be obtained by tracing back 
the radially outgoing null rays,  denoted by $H$ in Fig.1b.
As apparent from Fig.1, these can originate at the center of AdS prior to $p$.
In particular, provided the shell is sent in 
within time of order the AdS radius,
one can always make the resulting black hole large 
enough for its horizon to encompass the event $p$. 

\ifig\figb{Penrose diagram of the proposed process of abstracting information 
from inside of the horizon corresponding to the sketch in Fig.1b.}
{\epsfxsize=4cm \epsfysize=6.4cm \epsfbox{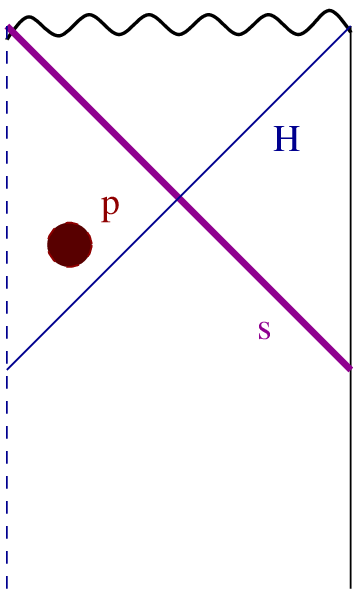}}

This implies that one has in fact succeeded in ``measuring'' an event,
$p$, which is inside a black hole horizon.
This is seen more clearly from the corresponding Penrose diagram, Fig.2.

\newsec{Discussion}

Several comments about this process are in order.
First, while we claim that, through precursors,
the boundary field theory encodes some information about events in
the interior of black holes, our claim does not imply violations of causality.
This is simply because the ``measurement'', $W$, is not local.  No message
is being sent between any two spacelike-separated points.
Correspondingly, one might then object that 1) a local field theory observer
can't set up the collapse of the shell fast enough if he waits till after
$W$ to decide whether to collapse the shell or not, and 2) a single local 
observer cannot get any information out of the nonlocal operator $W$ before
the shell is to be sent in.
Both of these statements are true, but they do not contradict the claim that
there is some quantity, $W$, which encodes some information about $p$, and 
yet is unaffected (due to the causality of the local quantum field theory 
on the boundary) by the eventual creation of the horizon.
In fact, this is suggestive of the generic expectation that extracting 
information out of black holes must necessarily involve some nonlocal 
physics.

Another objection that one might raise concerns the sudden creation of the
shell's energy, i.e.\ that there is a source introduced 
on the boundary.  
This however does not refute the basic point that 
$p$ can be measured despite being inside the horizon.
One may alternately set up the experiment with the shell's energy
being present for all time, as long as the shell implodes and creates
a black hole when required.
We have been talking about a ``shell'' without specifying how it
 is to be constructed from string theory, because such
detailed information is also irrelevant to the basic point.  
One can in fact consider more complicated nonspherical geometries, such as 
several colliding gravitational shock waves,
which may nevertheless be easier to discuss in
the dual field theory \hoit.

Clearly, for obtaining a specific demonstration of extracting 
information from a black hole,
 we would first need to specify exactly all the components
in our set-up before any explicit calculation could be carried out.
So far, the main obstacle to verifying our claim by an
explicit computation  is the incomplete
understanding of the holographic encoding of the bulk, namely the
exact nature of $W$.
A possible simplification is to work in lower dimensions
where exact solutions are easier to find, such 
as in $AdS_3 \times S^3 \times T^4$; however, even in this context, 
sufficient understanding of the precursors has not yet been achieved.

It is not  clear from our construction
 how much information one can actually extract out of 
the black hole, and what part of the spacetime inside the black hole 
can one extract information about.
In the above example, we claimed that we can learn something about
the part of  spacetime to the past of the shell, since there the 
picture is the same as for pure AdS.  
In terms of Fig.2, our process can  access the left wedge, 
but it is less clear whether we can similarly extract any information
from the top wedge.
Since the left wedge  is still far form the singularity, one might
worry that our proposal cannot hope address the  motivation
of resolving the singularity.
This does not appear to be a fundamental obstacle, since one may hope
to get closer to the singularity 
 by considering different types of collapse.

In summary,
we are still faced with the outstanding question of whether or not the gauge 
theory can probe physics inside the event horizon.
Since this region is causally disconnected from the boundary, 
one may fear that the existence of the horizon will prevent
the gauge theory from ``seeing'' what goes on behind it,
as is 
true for the local operators.
In this note 
we have shown that, given the existence of precursors, 
which are nonlocal operators, one can find situations
in which the boundary theory does know about events behind the horizon.
While our gedanken experiment does not address extracting
any information from an existing black hole,
it demonstrates the important point that
the causal properties
of the horizon do not pose a fundamental obstacle to learning something
about the physics behind it.
Once this crucial causal obstacle is removed, extracting {\it all}
information out of black holes may seem more hopeful.

\vskip 1cm

\centerline{\bf Acknowledgements}
I would like to thank
Lenny Susskind and Steve Shenker for encouraging me to write-up the idea,
and Gary Horowitz for initial discussions.
This work was supported by NSF Grant PHY-9870115.

%
%

\listrefs
\end